\documentclass[3p,authoryear]{elsarticle}
  \journal{European Journal of Control}

\usepackage{graphicx}
\usepackage{mathtools,amsmath,amsthm}
\usepackage{mathalfa}
\usepackage[bitstream-charter]{mathdesign}
\usepackage{tikz}
    \usetikzlibrary{calc}
    \usetikzlibrary{shapes.symbols}
   \usetikzlibrary{arrows.meta}
\usepackage{url}

\usepackage{enumitem}
\usepackage[english]{babel}
\usepackage{sublabel}
\usepackage{caption}
  
\newcommand\complex{{\mathbb C}}
\newcommand\myreal{{\mathbb R}}
\newcommand\NN{{\mathbb N}}
\newcommand{\diag}{\textrm{diag}}
\newcommand{\abs}[1]{\left|#1\right|}
\newcommand{\smat}[1]{\ensuremath{\left[ \begin{smallmatrix}#1\end{smallmatrix} \right]}}
\newcommand{\bmat}[1]{\ensuremath{\begin{bmatrix}#1\end{bmatrix}}}
\newcommand{\alphastar}{{\alpha^\star}}
\newcommand\ones{\ensuremath{{\mathbf 1}_N}}
\newcommand\zeros{\ensuremath{{\mathbf 0}}}

\newcommand{\A}{\ensuremath{\mathcal{A}}}
\newcommand{\Geq}{\ensuremath{\succeq}}
\newcommand{\Gst}{\ensuremath{\succ}}
\newcommand{\Leq}{\ensuremath{\preceq}}
\newcommand{\Lst}{\ensuremath{\prec}}
\newcommand{\overbar}[1]{\mkern 1.5mu\overline{\mkern-1.5mu#1\mkern+1.5mu}\mkern 1.5mu}
\newcommand{\adj}{{\mathcal W}}

\DeclareMathOperator{\e}{e}

\renewcommand{\Re}{\operatorname{Re}}
\renewcommand{\Im}{\operatorname{Im}}

\newtheorem{theorem}{Theorem}
\newtheorem{corollary}{Corollary}
\newtheorem{definition}{Definition}
\newtheorem{lemma}{Lemma}
\newtheorem{rremark}{Remark}
\newenvironment{remark}{\begin{rremark}\rm }{\hfill \hspace*{0.5pt}  \hfill $\lrcorner$\end{rremark}}

\begin{document}
\begin{frontmatter}

\title{
   Equivalent Conditions for the Synchronization of Identical Linear Systems over Arbitrary Interconnections
   \tnoteref{anr}
}

\author[UPS]{Nicola Zaupa}\ead{nzaupa@laas.fr }
\author[UniTN]{Giulia Giordano}\ead{giulia.giordano@unitn.it}
\author[Laas]{Isabelle Queinnec}\ead{queinnec@laas.fr }
\author[Laas]{Sophie Tarbouriech}\ead{tarbour@laas.fr }
\author[UniTN,Laas]{Luca Zaccarian}\ead{zaccarian@laas.fr }
\tnotetext[anr]{Work funded in part by the European Union under NextGenerationEU (PRIN 2022 project PRIDE, grant number 2022LP77J4, CUP E53D23000720006) and by the ANR via grant OLYMPIA, number ANR-23-CE48-0006.}
\tnotetext[]{Available in open access at \url{https://doi.org/10.1016/j.ejcon.2024.101099}.}
\address[UPS]{LAAS -- CNRS, Universit\'e de Toulouse, UPS, Toulouse, France.}
\address[UniTN]{Dipartimento di Ingegneria Industriale, Università di Trento, Trento, Italy.}
\address[Laas]{LAAS -- CNRS, Universit\'e de Toulouse, CNRS, Toulouse, France.}

\begin{abstract}
   We propose necessary and sufficient conditions for the synchronization of $N$ identical single-input-single-output (SISO) systems, connected through a directed graph {without imposing any assumption on the graph interconnection}.
   We consider both the continuous-time and the discrete-time case, and we provide conditions that {are equivalent to} the uniform global exponential stability, {with guaranteed convergence rate,} of the closed {and unbounded} attractor that corresponds to the synchronization set.
\end{abstract}


\end{frontmatter}

\section{Introduction}

   The problems of consensus and synchronization of multi-agent systems \citep{FagnaniBook2017} have received growing interest, due to the variety of applications
   in many different areas, including: cooperative control of unmanned aerial vehicles, formation control of mobile robots and communication in sensor networks
   \citep{fax2004, jadbabaie2003,ren2007}, quality-fair delivery of media contents \citep{dalcol2017}, power networks \citep{dorfler2013}, biological systems \citep{scardovi2010}, and opinion dynamics \citep{anderson2019}.
   Specifically, \textit{consensus} refers to agents coming to a global agreement on 
   a state value, thanks to the exchange of information modeled by some communication graph; mild assumptions on the graph connectivity allow to uniformly exponentially reach consensus \citep{jadbabaie2003, olfati-saber2004, olfati-saber2007,moreau2005,ren2008,wieland2008,seo2009}.
   Conversely, \textit{synchronization} refers to agents moving toward a common trajectory in the configuration space \citep{hale1997,pecora1998,slotine2005,scardovi2009,carli2011,sepulchre2011,andrieu2018,dalcol2018,dalcol2019}.
   Consensus and synchronization problems have been widely investigated for agents modeled by identical linear time-invariant (LTI) systems, with many subsequent extensions to switching network topologies \citep{olfati-saber2004,xiao2007,su2012}, heterogeneous and nonlinear systems \citep{khong2016,PL2019,AMP2021}, output synchronization \citep{zhu2016,IsidoriBook2017}, and hybrid systems \citep{mayhew2012,teel2015,cristofaro2022}.

\begin{figure}[bt]
      \centering
      \includegraphics[scale=0.7]{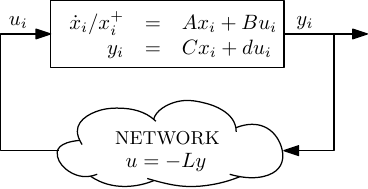}
      \caption{Block diagram of the closed-loop system.}
      \label{fig:network}
   \end{figure}

   Figure~\ref{fig:network} represents the distributed feedback system addressed in this work, where $N$ identical SISO dynamical systems of arbitrary order, with state $x_i \in \mathbb{R}^n$ evolving as
   \begin{equation}\label{eq:sys}
      \begin{array}{rcl}
      \dot {x}_i/x_i^+ & =& A x_i + B u_i \\
      y_i              & =& C x_i + d u_i \\
      \end{array} \quad\quad i=1, \dots, N
   \end{equation}
   and with scalar inputs $u_i \in \mathbb{R}$ and outputs $y_i \in \mathbb{R}$,
   are interconnected through a directed graph  
   $\mathcal G$ with Laplacian $L\in\myreal^{N\times N}$ as follows:
   \begin{equation} \label{eq:laplacian}
   u = -L y = -L\left(  (I_N \otimes C)x + du \right),  
   \end{equation}
   where  $u =  \left[ u_1 \dots u_N \right]^{\top} \in \mathbb{R}^{N}$,
   $y = \left[ y_1 \dots y_N \right]^{\top}\in \mathbb{R}^{N}$ and
   $x := \left[x_1^{\top} \dots x_N^{\top} \right]^{\top} \in \mathbb{R}^{N n}$,
   are the aggregate input, output and state vectors, respectively, and $\otimes$ denotes the Kronecker product. {No explicit assumptions are imposed a priori on the connectivity properties {of the graph that describes} the interconnection network.}   

   Due to linearity, a range of equivalent conditions for synchronization can be stated and may lead to powerful numerical tools for distributed controllers tuning.
   Despite the extensive amount of work in the field, one cannot find a \textit{general} theorem clearly stating the equivalence of these conditions for
   identical LTI systems of \textit{arbitrary} order, connected through an \textit{arbitrary} graph topology. 
   Here, we provide this result by introducing a list of necessary and sufficient conditions for uniform global exponential synchronization with guaranteed convergence rate both in the continuous-time and in the  discrete-time case.
   For the sake of generality, we also allow for the presence of a direct input-output link $d\in\myreal$ in \eqref{eq:sys}.
   With $d\neq 0$, the implicit equation \eqref{eq:laplacian} can be uniquely solved as
   \begin{equation}
         u = -({L_d} \otimes C) x := -((I_N+dL)^{-1}L \otimes C)x 
   \end{equation}
   if and only if (linear) well-posedness holds, namely matrix \mbox{$I_N+dL$} is invertible or, equivalently, $-d^{-1}$ is not an eigenvalue of $L$.
   Clearly, when $d=0$ we retrieve the classical case
   ${L_d} := (I_N+dL)^{-1}L = L$, but our results also characterize the general case with a distributed algebraic loop.
   Overall, the linear distributed interconnection \eqref{eq:sys}, \eqref{eq:laplacian}
   can be  written as
   \begin{equation}\label{eq:closedloop}
      \dot x {/ x^+} = (I_N \otimes A) x - ({L_d} \otimes BC)x.
   \end{equation}

   For dynamics \eqref{eq:closedloop}, we focus on {the synchronization set $\A$}, i.e., the (unbounded) set where all pairwise states coincide. Then, our main contribution consists in providing a list of necessary and sufficient conditions for the uniform global exponential stability of {the} synchronization set {$\A$} with convergence rate guarantees.
   Our necessary and sufficient conditions comprise 
   (a) Hurwitz/Schur properties of {certain} complex-valued matrices induced by the eigenvalues of $L_d$, 
   (b) equivalent Hurwitz/Schur properties of suitable real-valued matrices, 
   (c) existence of positive-definite solutions to certain Lyapunov inequalities, 
   (d) existence of a strict quadratic Lyapunov function (establishing a quadratic converse Lyapunov theorem), and 
   (e) synchronization of all solutions towards a specific initial value problem.

   Proving equivalence of the above properties is a contribution per se, as typically one finds only parts of these equivalences in the literature, possibly with different assumptions on the Laplacian $L$.
   Moreover, for the construction of the strict quadratic Lyapunov function in item (d) above, we adopt a scaling method issued from stability of cascaded systems, which generalizes the typical block-diagonal Lyapunov functions comprising multiple identical block-diagonal components.
   Conditions of the same form as (a) for formation stability were given in \citep[Theorem~3]{fax2004} and the uniform global exponential stability condition was exploited in \citep[Theorem~1]{scardovi2014consensus} and \citep[Theorem~1]{seo2009}.
   The condition related to the initial value problem (e) was given in \citep{scardovi2009}.
   Moreover, a list of similar discrete-time statements was given under an undirected graph assumption in \citep[Theorem~1]{dalcol2017}.

   Our conditions do not require any assumptions on the graph connectivity properties. Still, if we require the synchronization set to be a \textit{non-trivial} solution, we can consider connected graphs without loss of generality; in fact, for a disconnected graph, synchronization can be achieved (without information exchange) only if it is trivial (i.e., all solutions converge to the origin).

   The presented necessary and sufficient conditions could be used to parameterize all possible stabilizers and possibly select the \emph{best} one from a certain performance view-point;
   an example of this approach is provided in the recent work \citep{zaupa2023}, where analogous conditions are used for the design of a simultaneous stabilizing static state-feedback controller, which is then compared with existing approaches where Riccati-based designs are used to show the existence of a stabilizer.
   Moreover, our proposed conditions can be used to extend the LMI-based $H_\infty$ results in \citep{dalcolThesis2016}, as well as the LMI-based saturated feedback design in \citep{dalcol2019}, to also deal with undirected graphs. 
   More in general, while existence results for global simultaneous stabilizers are already 
   available from \citep{IsidoriBook2017,SaberiBook2022}, 
   our conditions become relevant when solving a multi-objective control (or observer) design problem.

   The paper is structured as follows. Section~\ref{sec:mainresult} presents basic definitions along with the main theorem, whose proof is provided in Section~\ref{sec:proof}. 

\smallskip

\noindent
{\bf Notation.}  
$\myreal$ ($\myreal_{\geq0}$), $\mathbb{C}$ and $\mathbb N$ denote respectively the sets of real (non-negative), complex and natural numbers. 
We denote with $\jmath$ the imaginary unit and with $\ones \in \mathbb{R}^N$ ($\mathbf 0_N \in \mathbb{R}^N$) the $N$-dimensional (column) vector having all $1$ entries ($0$ entries);
$\diag \left(A_1, \dots, A_N \right)$
indicates the block-diagonal matrix whose  diagonal
blocks are the square matrices $A_1, \dots, A_N$.
Given a complex number $\lambda = a+\jmath b$, $\Re(\lambda)=a$ denotes its real part, $\Im(\lambda) = b$ its imaginary part, and $\lambda^* = a-\jmath b$ its complex conjugate.
Given a complex matrix $A \in \complex^{n\times m}$, $A^*$ denotes its conjugate transpose.
A square matrix $A \in \complex^{n \times n}$ is Hurwitz when all its eigenvalues have strictly negative real part; 
it is Schur when all its eigenvalues have modulus strictly less than $1$;
it is Hermitian if $A = A^*$, namely $\Re(A)$ is symmetric $\left(\Re(A)=\Re(A)^\top\right)$ and $\Im(A)$ is skew-symmetric $\left(\Im(A)=-\Im(A)^\top\right)$. 
A Hermitian matrix $A$ has real eigenvalues and is positive semi-definite, $A\Geq 0$ (resp. positive definite, $A\Gst 0$), when
its eigenvalues are all non-negative (resp. positive).
We denote with $\sigma(A)$ the spectrum of a square matrix ${A}$, and we call \textit{dominant} the eigenvalue having the largest real part.
Eigenpairs for which $Av=\lambda v$ are denoted as $(\lambda,v)$.
We denote the Euclidean distance of a point $x$ from a set ${\mathcal {H}}$ as $|x|_{\mathcal H}:= \inf\limits_{y\in \mathcal H} \abs{x-y}$.
The topology of a directed graph $\mathcal G$ with $N \in {\mathbb N}$ nodes  is characterized by the weighted adjacency matrix $\adj \in \mathbb{R}^{N \times N}$ whose entry $\mathcal W_{ij}\geq 0$ denotes the weight of the edge pointing from node $j$ to node $i$. Defining the diagonal matrix $D:= \diag(\adj \ones)$, we can introduce the Laplacian matrix $L := D-\adj$ associated with the graph 
$\mathcal G$. 


\section{Equivalent conditions to $\alpha$--synchronization}
\label{sec:mainresult}

   To establish synchronization among systems (\ref{eq:sys}), 
   interconnected via \eqref{eq:laplacian}, as represented in Figure~\ref{fig:network},
   we first introduce the synchronization set, or {\em attractor},
   \begin{equation}\label{eq:setA}
   \A: = \begin{Bmatrix} x  : \, x_i - x_j = 0, \, 
   \forall i,j \in  \left \{ 1, \dots , N \right \} \end{Bmatrix}
   \end{equation}
   along with the definitions of exponential stability and synchronization for the continuous-time and the discrete-time case, for a given convergence rate $\alpha$.
   \begin{definition}[$\alpha$--UGES] \label{def:alpha_stability}
      The attractor $\A$ in \eqref{eq:setA} is $\alpha$-UGES (uniformly globally exponentially stable with rate $\alpha>0$) for system \eqref{eq:sys}, \eqref{eq:laplacian} if there exists $M>0$ such that any solution $t\mapsto x(t)$ satisfies
      \begin{subequations}
         \begin{equation}
            |x(t)|_\A \leq M\e^{-\alpha t}|x(0)|_\A,\quad \forall t\in\myreal_{\geq0}
         \end{equation}
         \begin{equation}
            \bigl( resp.\ |x(t)|_\A \leq M\alpha^{t}|x(0)|_\A,\quad \forall t\in\NN \bigr).
         \end{equation}
      \end{subequations}
   \end{definition}

   \begin{remark} \label{remark:exp_asym}
      The definition above considers exponential stability of the attractor, which coincides with its asymptotic stability due to the linear and homogeneous dynamics of the system.
   \end{remark}

   \begin{definition}[$\alpha$--synchronization] \label{def:alpha_synch}
      For the continuous-time (respectively discrete-time) linear system \eqref{eq:sys}, \eqref{eq:laplacian}, 
      $\alpha$--synchronization holds if there exist $M>0$ and rate $\alpha>0$ (resp. $\alpha\in(0,1)$) such that, for any initial condition, every sub-system satisfies
      \begin{subequations}
      \begin{equation}
         |x_i(t)-{\tilde x_\circ(t)}| \leq M\e^{-\alpha t} {\sum_{i=1}^N}|x_i(0)-{\tilde x_\circ(0)}|,\quad\forall t\in\myreal_{\geq0}
      \end{equation}
      \begin{equation}
         \bigl( resp.\ |x_i(t)-{\tilde x_\circ(t)}| \leq M\alpha^{t} {\sum_{i=1}^N}|x_i(0)-{\tilde x_\circ(0)}|,\quad\forall t\in\NN 
         \bigr) ,
      \end{equation}
      \end{subequations}
      $i=1,\ldots,N$, for a suitable target trajectory ${\tilde x_\circ}$, depending on the initial conditions.
   \end{definition}


   Before stating our main result, 
   we note that, trivially, all the zero eigenvalues of $L$ are also zero eigenvalues of $L_d = (I_N+dL)^{-1}L$. Then
   we denote by $0=\lambda_0, \lambda_1, \dots,\lambda_\nu$ the eigenvalues of $L_d$, where the complex conjugate pairs are only counted once  (so that $\nu := N- 1-n_c$, with $n_c$ being the number of complex conjugate pairs).

   Given an assigned convergence rate $\alphastar \geq 0$ of the solutions towards the attractor $\A$, we state a list of necessary and sufficient conditions for (continuous- or discrete-time) $\alphastar$--exponential synchronization of (\ref{eq:sys}), \eqref{eq:laplacian}. 

\begin{theorem}\label{th:consensus_alpha}
   Consider the continuous-time (resp. discrete-time) system in \eqref{eq:sys}, \eqref{eq:laplacian}, the attractor 
   $\A$ in \eqref{eq:setA} and the parameter $\alphastar\geq0$ (resp. $\alphastar\in(0,1]$).
   The following statements are equivalent:
   \begin{enumerate}
   \item \label{it:Ak_alpha} \emph{[Complex condition]}
      The spectral abscissa (resp. spectral radius) of the complex-valued matrices\footnote{With complex conjugate pairs $\lambda_k,\ \lambda_k^*$,
      it is enough to check condition \eqref{eq:Akdef} for one of the two eigenvalues.
      Indeed, if $(A-\lambda_k BC)(v+\jmath w) = \mu (v+\jmath w)$, then by taking the conjugate we get 
      $(A-\lambda_k^* BC)(v-\jmath w) = \mu^* (v-\jmath w)$, 
      where $\mu$ and its conjugate $\mu^*$ have the same real part and the same modulus.}
      \begin{align} \label{eq:Akdef}
         A_k &:= A - \lambda_k BC ,  \quad  k=1, \dots, \nu ,
      \end{align}
      is smaller than $-\alphastar$ (resp. $\alphastar$).
   \item \label{it:Akreal_alpha} \emph{[Real condition]} 
   The spectral abscissa (resp. spectral radius) of the real-valued matrices
      \begin{equation}\label{eq:Aekdef} 
         A_{e,k} := \begin{bmatrix} A - \Re (\lambda_k) BC & \Im(\lambda_k) BC \\
         - \Im(\lambda_k) BC & A - \Re (\lambda_k) BC 
         \end{bmatrix},   \quad  k=1, \dots, \nu ,
      \end{equation}
      is smaller than $-\alphastar$ (resp. $\alphastar$).
   \item \label{it:lyapReal_alpha} \emph{[Lyapunov inequality]}
   For each $k=1, \dots, \nu$, there exist real-valued matrices $P_k = P_k^\top \Gst 0$ and $\Pi_k^\top = -\Pi_k$
   such that one of the following identities holds:
   \begin{subequations}
      \label{eq:Lyap_alpha_cmpx}
      \begin{equation}
         (P_k+\jmath \Pi_k) A_{k} + A_{k}^* (P_k+\jmath \Pi_k) \Lst -2\alphastar (P_k+\jmath \Pi_k)
      \end{equation}
      \begin{equation}
         \bigl(\ \text{resp.}\quad A_{k}^* (P_k+\jmath \Pi_k) A_{k}  \Lst (\alphastar)^2 (P_k+\jmath \Pi_k) \ \bigr),
      \end{equation}
   \end{subequations}
   or
   \begin{subequations}
      \label{eq:Lyap_alpha_real}
      \begin{equation}
         \bmat{P_k  &  \Pi_k \\ -\Pi_k & P_k} A_{e,k} + A_{e,k}^\top \bmat{P_k  &  \Pi_k \\ -\Pi_k & P_k} \Lst -2\alphastar \bmat{P_k  &  \Pi_k \\ -\Pi_k & P_k}
      \end{equation}
      \begin{equation}
         \left(\ \text{resp.}\quad A_{e,k}^\top \bmat{P_k  &  \Pi_k \\ -\Pi_k & P_k} A_{e,k}   \Lst (\alphastar)^2 \bmat{P_k  &  \Pi_k \\ -\Pi_k & P_k} \ \right).
      \end{equation}
   \end{subequations}
   \item \label{it:lyap_alpha}   \emph{[Lyapunov function]}
      There exist $\alpha>\alphastar$ (resp. $0<\alpha<\alphastar$),
      positive constants $c_1$, $c_2$ and $c_3$, 
      and a strict quadratic Lyapunov function $V(x)$ satisfying:
      \begin{subequations} \label{eq:V_lyap}
         \begin{align} 
         \label{eq:V_lyap_sandwich}
            & c_1 \abs{x}_{\A}^2 \leq V(x) \leq  c_2 \abs{x}_{\A}^2,\\
         \label{eq:V_lyap_var_CT}
            & \dot{V}(x)  \leq - 2\alpha V(x)\\
         \label{eq:V_lyap_var_DT}
             \bigl( \text{resp.}\  &V(x^+)   \leq \alpha^2 V(x) \ \bigr).
         \end{align}
      \end{subequations}
   \item \label{it:UGES_alpha} \emph{[UGES]}
      There exists $\alpha>\alphastar$ (resp.  $0<\alpha<\alphastar$) such that the closed attractor $\A$ in (\ref{eq:setA})
      is  $\alpha$--UGES
      for the closed loop (\ref{eq:sys}), (\ref{eq:laplacian}).
   \item \label{it:attractor_alpha} \emph{[IVP]}
      There exists $\alpha>\alphastar$ (resp. $0<\alpha<\alphastar$) such that, for the closed loop (\ref{eq:sys}), (\ref{eq:laplacian}),
      the sub-states $x_i$ uniformly globally $\alpha$--exponentially synchronize to the unique solution of the following initial value problem:
      \begin{equation}\label{eq:xzero}
         \dot{\tilde {x}}_{\circ} / \tilde {x}^+_{\circ} = A  \tilde x_{\circ},  
         \quad \tilde x_{\circ} (0) = \frac{1}{p^\top{\bf 1}_N} \sum\limits_{k = 1}^{N} p_k x_k(0),
      \end{equation}
      where $p:=[p_1  \dots  p_N]^\top \in \mathbb{R}^N$ is a left eigenvector corresponding to the zero eigenvalue of $L$.
   \end{enumerate}
\end{theorem}

\begin{remark}
   Referring to item~\ref{it:attractor_alpha} above,
   the $k$-th entry of the left eigenvector $p$, associated with the zero eigenvalue of the Laplacian $L$, can be seen as a measure of the \textit{centrality} of node $k$, which weighs its initial state in the linear combination $\tilde x_o(0)$ in \eqref{eq:xzero}; a strong analogy can be observed with the Bonacich centrality \citep{Bonacich1987}, which is the left eigenvector associated with the $1$ eigenvalue of the normalized adjacency matrix.
\end{remark}

\begin{remark}
   Theorem~\ref{th:consensus_alpha} holds also for weakly-connected or even disconnected graph. 
   In this situation, the algebraic multiplicity of the zero eigenvalue of $L$ is greater than one and therefore we have the trivial condition that matrix $A$ has to be Hurwitz/Schur.
   This is a reasonable conclusion: for a disconnected network, the only possible common equilibrium without information exchange is the origin.
   Moreover, referring to item~\ref{it:attractor_alpha}, in this case the vector $p$ is not uniquely determined (up to rescaling) since there exist as many linearly independent eigenvectors as  the geometric multiplicity of the zero eigenvalue. 
   In fact, any such selction of $p$ is a valid one for item~\ref{it:attractor_alpha} because, with disconnected networks, all the equivalent items of the theorem are true if and only if the solution of \eqref{eq:xzero} converges to zero (a trivial synchronized motion).
\end{remark}


In the special case where $\alphastar=0$ (resp. $\alphastar=1$) and \mbox{$d=0$}, matrix $L_d$ becomes the Laplacian $L$ of the graph, and the results in Theorem~\ref{th:consensus_alpha} can be stated in a simplified (and widely studied) setting, as clarified in the next corollary.

\begin{corollary} \label{cor:main}
   Consider the continuous-time (resp. discrete-time) system \eqref{eq:sys}, \eqref{eq:laplacian} with $d=0$, so that  $0=\lambda_0, \lambda_1, \dots,\lambda_\nu$ are the eigenvalues of $L$.
   The following statements are equivalent:
   \begin{enumerate}
   \item  \emph{[Complex condition]}
   The complex-valued matrices \eqref{eq:Akdef} are Hurwitz (resp. are Schur).
   \item  \emph{[Real condition]} 
   The real-valued matrices \eqref{eq:Aekdef} are Hurwitz (resp. are Schur).
   \item  \emph{[Lyapunov inequality]}
   For each $k=1, \dots, \nu$, there exist real-valued matrices $P_k = P_k^\top \Gst 0$ and $\Pi_k^\top = -\Pi_k$
   such that one of the following identities holds:
   \begin{equation*}
      (P_k+\jmath \Pi_k) A_{k} + A_{k}^* (P_k+\jmath \Pi_k) \Lst 0 \quad
      \bigl(\ \text{resp.}\quad A_{k}^* (P_k+\jmath \Pi_k) A_{k}  \Lst 0 \ \bigr),
   \end{equation*}
   or
   \begin{equation*}
      \bmat{P_k  &  \Pi_k \\ -\Pi_k & P_k} A_{e,k} + A_{e,k}^\top \bmat{P_k  &  \Pi_k \\ -\Pi_k & P_k} \Lst 0 \quad
         \left(\ \text{resp.}\quad A_{e,k}^\top \bmat{P_k  &  \Pi_k \\ -\Pi_k & P_k} A_{e,k}   \Lst 0 \ \right).
   \end{equation*}
   \item  \emph{[Lyapunov function]}
   There exist positive constants $c_1$, $c_2$ and $c_3$
   and a strict quadratic Lyapunov function $V$ satisfying \eqref{eq:V_lyap_sandwich} and \mbox{$\dot{V}(x)  \leq -c_3V(x)$} 
   (resp. \mbox{$V(x^+)   \leq (1-c_3)V(x)$}) for all $x\in \myreal^{Nn}$.
   \item  \emph{[UGES]}
   The closed attractor $\A$ in \eqref{eq:setA}
   is  uniformly globally exponentially stable.
   \item  \emph{[IVP]}
   The closed loop \eqref{eq:sys}, \eqref{eq:laplacian}  is such that the 
   sub-states $x_i$ uniformly globally exponentially synchronize to the unique 
   solution of the initial value problem \eqref{eq:xzero}.
   \end{enumerate}
\end{corollary}

\begin{remark} 
   In the continuous-time case, following \citep[Lemma 1]{hara2014}, comparing the dynamic matrix ${\mathbf A}:=( I_N \otimes A) -( L \otimes BC)$ arising from \eqref{eq:closedloop} with \citep[equation (4)]{hara2014}, we may give an additional frequency-domain condition, equivalent to the above items, expressed in terms of a coprime factorization $\frac{\overbar n(s)}{\overbar d(s)} = C(sI-A)^{-1}B$ of dynamics \eqref{eq:sys} as follows:

   \smallskip

   \noindent
   \textit{(vii). \emph{[Frequency domain]} $\sigma(L) \subset \Lambda:=\{\lambda\in \complex: \overbar d(s)+\lambda \overbar n(s) 
   \mbox{ is Hurwitz} \}$.}
\end{remark}

\section{Proof of the main theorem}
\label{sec:proof}

\subsection{A few technical lemmas}
   Before proving Theorem~\ref{th:consensus_alpha}, we state some preliminary facts useful for the proof.
   We prove the following standard result to highlight that no assumptions on the graph $\mathcal G$ are needed.

   \begin{lemma} \label{lem:Laplacian_eig}
      Given any (directed) graph $\mathcal G$ and its Laplacian $L= D-\adj$, the eigenvalues of $M :=-L$ have non-positive real part and the dominant eigenvalue of $M$ is $\mu_0 = 0$.
      It is associated with a right eigenvector $\ones$ and a left eigenvector $p \in \mathbb{R}^N$ that can be selected non-negative.
   \end{lemma}

   \begin{proof}
      Since $\adj$ has non-negative elements by construction and $D$ is diagonal, $M = \adj-D$ is a Metzler matrix (i.e., its off-diagonal entries are non-negative). 
      Hence, as a consequence of Perron-Frobenius theory,  the dominant eigenvalue $\mu_0$ of $M$ is real and associated with left and right eigenvectors having non-negative elements \citep[Chapter 6.5, Theorem 1]{Luenberger}.
      In view of Gershgorin's Circle Theorem \citep[Chapter 6.1]{HornBook2012}, the eigenvalues of $M$ lie in the union of the $N$ disks centered at $-D_{ii}$ and with radius $\sum_{j=1}^N \adj_{ij}=D_{ii}$, $i=1,\dots,N$, which are all included in the left half plane and tangent to the imaginary axis at zero. Therefore, all the eigenvalues have non-positive real part and the only possible eigenvalue whose real part is not strictly negative is zero. $M$ is singular, because $M \ones = \zeros_N$. 
      Then, $\mu_0 = 0$ is the dominant eigenvalue and we can select $\ones$ as a right eigenvector and a vector with non-negative elements as a left eigenvector.
   \end{proof}
   
   We will rely on the following lemma for the decomposition of the closed-loop dynamics.
   
   \begin{lemma} \label{lem:L_frob}
      Consider the matrix ${L_d} \in \mathbb{R}^{N \times N }$ defined as ${L_d}=(I_N+dL)^{-1}L$, where $L \in \mathbb{R}^{N \times N }$ is the Laplacian of a directed graph $\mathcal G$ and $d\in\myreal$ is such that $(I_N+dL)$ is invertible.
      There exists an orthogonal matrix $T$, whose first column is $\frac{1}{\sqrt{N}} \ones$,   
      that transforms $L_d$ into an upper block-triangular matrix:
      \begin{equation}
      \label{eq: L_bar_frob}
         \overbar L_d := T^\top  L_d T = 
         \smat{
         \lambda_0 & \star      & \cdots  & \star  \\
                  0 & \Lambda_1  & \cdots  & \star  \\
            \vdots & \vdots     & \ddots & \vdots \\
                  0 & 0          & 0      & \Lambda_{\nu}
         } = \smat{
               0 & \star     & \cdots  & \star  \\
               0 & \Lambda_1 & \cdots  & \star  \\
         \vdots & \vdots    & \ddots & \vdots \\
               0 & 0         &  0     & \Lambda_{\nu}
         },
      \end{equation}
      where blocks $\Lambda_i$, $i=1,\dots,\nu$, are either scalar (corresponding to real eigenvalues of $L_d$) or 2-by-2
      matrices (corresponding to complex conjugate eigenvalue pairs of $L_d$).
   \end{lemma}

\begin{proof}
   First, from \citep[Theorem 2.3.4, item (b)]{HornBook2012} it follows that there exists an orthogonal $T$ such that $L_d$ can be decomposed as in \eqref{eq: L_bar_frob} with no particular structure for $\Lambda_i$.
   Second, $L$ and ${L_d}$ share the same eigenspace associated with the zero eigenvalue. 
   To show this, let us consider the eigenpair $(0,w)$ for $L$, such that $Lw=0$. Then, since ${L_d}=(I_N+dL)^{-1}L$, we have that ${L_d}w=(I_N+dL)^{-1}Lw=0$.
   This means that $(0,w)$ is also an eigenpair for ${L_d}$ and therefore the zero eigenvalue has the same algebraic and geometric multiplicity for both $L$ and $L_d$.
   Third, from \citep[Theorem 4]{agaev2005} and \citep[Corollary 4.2]{CaughmanVeerman2006}, the zero eigenvalue of $L$ is semisimple (i.e., its algebraic and geometric multiplicities coincide), which holds also for $L_d$ since they share the same eigenspace.
   Finally, in order to have the first entry of $\overbar L_d$ equal to zero and $T$ orthogonal, the first column of $T$ must be equal to $\frac{1}{\sqrt{N}} \ones$ so that $\left(\frac{1}{\sqrt{N}} \ones\right)^\top\left(\frac{1}{\sqrt{N}} \ones\right)=1$. 
\end{proof}

\begin{remark}
In our proofs (and in particular in Lemma~\ref{lem:L_frob}), we do not assume that the zero eigenvalue of $L_d$ is simple, but we leverage the fact that, for any Laplacian matrix, zero is a semisimple eigenvalue, without the need to assume connectedness or any other property of the graph topology. At the same time, 
items (i) and (ii) of Corollary~\ref{cor:main} clearly indicate that -- excluding the trivial case where $A$ is Hurwitz (resp. Schur), and all trajectories synchronize to the trivial solution -- connectedness of the graph (namely, having a nonzero second eigenvalue of $L$) is a necessary condition for synchronization. While this is a well-known fact, we emphasize that in general connectedness is not enough to establish synchronization and sufficient conditions often require a lower bound on the second-smallest eigenvalue. In this sense our conditions may generally lead to reduced conservativeness, as confirmed by their necessity.   
\end{remark}

We will also use the following results, which trivially follow from \citep{stykel2002} for the particular case $E=I$, that guarantee necessary and sufficient conditions for a complex matrix  to be Hurwitz or Schur. 
In particular, for the continuous-time case we consider \citep[Theorem~2.3]{stykel2002}.

\begin{lemma} \label{lem:scardovi}
   A matrix $S\in \complex^{n \times n}$ is Hurwitz if and only if, for each positive definite $Q\in \complex^{n \times n}$, $Q^* = Q$, there exists a positive definite $H\in \complex^{n \times n}$, with $H^* = H$, such that
   $S^* H + HS =-Q$.
\end{lemma}

For the discrete-time case we consider \citep[Theorem~3.2]{stykel2002}, which is a generalization of the results in \citep[Theorem~7]{wimmer1973}.

\begin{lemma} \label{lem:schur_complex}
   A matrix $S\in \complex^{n \times n}$ is Schur if and only if, for each positive definite $Q\in \complex^{n \times n}$, $Q^* = Q$, there exists a positive definite $H\in \complex^{n \times n}$, with $H^* = H$, such that
   $S^* HS - H =-Q$.
\end{lemma}

   For the continuous-time case, the proof combines the stability results in \citep{fax2004} 
   with the  output feedback coupling approach of \citep{scardovi2009}.  
   Parts of the result can be found in the literature,
   possibly with different assumptions on the Laplacian $L$.
   For example, 
   necessary and sufficient conditions of the form \ref{it:Ak_alpha} for formation 
   stability were given in \citep[Theorem 3]{fax2004};
   implication  
   \mbox{\ref{it:Ak_alpha}   $\implies$  \ref{it:UGES_alpha}}
   was established in an equivalent formulation in \citep[Theorem 1]{scardovi2014consensus} and \citep[Theorem 1]{seo2009} 
   for the convergence part. 
   In \citep{wieland2011}, the equivalence \ref{it:Ak_alpha} $\iff$ \ref{it:attractor_alpha} 
   was proven.
   For the discrete-time case, in \citep[Theorem 1]{dalcol2017} a similar statement is given under an undirected graph assumption.   

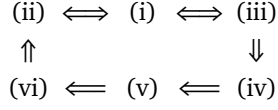
\begin{figure}
    \centering
    \begin{tikzpicture}
        \draw (0  ,0) node(A){\ref{it:Akreal_alpha}} 
            ++(1.5,0) node(B){\ref{it:Ak_alpha}} 
            ++(1.5,0) node(C){\ref{it:lyapReal_alpha}};
        \draw (C)++(0,-1) node(D){\ref{it:lyap_alpha}}
            ++(-1.5,0) node(E){\ref{it:UGES_alpha}}
            ++(-1.5,0) node(F){\ref{it:attractor_alpha}} ;
        \draw ($(A)!0.5!(B)$) node{$\Longleftrightarrow$};
        \draw ($(B)!0.5!(C)$) node{$\Longleftrightarrow$};
        \draw ($(C)!0.5!(D)$) node{$\Downarrow$};
        \draw ($(F)!0.5!(A)$) node{$\Uparrow$};
        \draw ($(D)!0.5!(E)$) node{$\Longleftarrow$};
        \draw ($(E)!0.5!(F)$) node{$\Longleftarrow$};
    \end{tikzpicture}
    \caption{Structure of the proof of Theorem~\refeq{th:consensus_alpha}.}
    \label{fig:proofstructure}
\end{figure}

\subsection{Proof of Theorem~\refeq{th:consensus_alpha}}

   To show the equivalence of the six statements in Theorem~\refeq{th:consensus_alpha}, the proof is structured as follows. We first prove the equivalence among statements \ref{it:Ak_alpha}, \ref{it:Akreal_alpha} and \ref{it:lyapReal_alpha}. Then, we prove the following chain of implications:
   \ref{it:lyapReal_alpha} $\Longrightarrow$ \ref{it:lyap_alpha}, followed by \ref{it:lyap_alpha} $\implies$ \ref{it:UGES_alpha}, \ref{it:UGES_alpha} $\implies$ \ref{it:attractor_alpha},  and finally \ref{it:attractor_alpha} $\implies$ \ref{it:Akreal_alpha}. This ensures that all the six statements are equivalent.
   A pictorial representation of the above-mentioned proof structure is illustrated in Figure~\ref{fig:proofstructure}.

   \smallskip

   \noindent
   {\bf\em Proof of \ref{it:Ak_alpha} $\Longleftrightarrow$ \ref{it:Akreal_alpha}}. 
   Given the complex-valued matrix $A_k = M_k + \jmath N_k \in \complex^{n \times n}$, where $M_k, N_k \in \myreal^{n \times n}$, following \citep[Fact~2.17.3]{bernstein2005MatrixMathematics} we can write
   \begin{equation}
      \label{eq:luca}
      A_{e,k} = \bmat{M_k & N_k\\ -N_k & M_k } = \bmat{I_n & 0 \\ \jmath I_n & I_n}\bmat{M_k+\jmath N_k & N_k \\ 0 & M_k-\jmath N_k} \bmat{I_n & 0 \\ -\jmath I_n & I_n}, 
   \end{equation}
   from which it follows that $\sigma(A_{e,k})=\sigma(A_k)\cup\sigma(A_k^*)$.
   The implication \ref{it:Ak_alpha} $\Longleftarrow$ \ref{it:Akreal_alpha} directly follows from the fact that \eqref{eq:luca} implies $\sigma(A_k)\subseteq\sigma(A_{e,k})$.
   To show \ref{it:Ak_alpha} $\Longrightarrow$ \ref{it:Akreal_alpha}, note that $\sigma(A_k^*)=(\sigma(A_k))^*$, which means that the spectral abscissa (resp. the spectral radius) of $A_k$ and $A_k^*$ are the same.
   As a consequence, also the union of their spectrum preserves the same properties, i.e., $A_{e,k}$ has the same spectral abscissa (resp. the same spectral radius) as $A_k$, thus concluding the proof of \ref{it:Ak_alpha} $\Longleftrightarrow$ \ref{it:Akreal_alpha}.

   \smallskip

   \noindent
   {\bf\em Proof of \ref{it:Ak_alpha} $\Longleftrightarrow$ \ref{it:lyapReal_alpha}}. 
      Statement \ref{it:Ak_alpha} is equivalent to imposing that $A_k+\alphastar I$ is Hurwitz (resp. $\frac{1}{\alphastar}A_k$ is Schur).
      Therefore, for the continuous-time case, in view of Lemma \ref{lem:scardovi}, Hurwitz stability of $A_k+\alphastar I$ is equivalent to $A_k^* H_k + H_k A_k \Lst -2\alphastar H_k$ for a positive definite Hermitian matrix $H_k=H_k^*$.
      By splitting real and imaginary part, we write $A_k = M_k + \jmath N_k$ and $H_k = P_k + \jmath \Pi_k$, where $P_k=P_k^\top$ and $\Pi_k = -\Pi_k^\top$ because $H_k$ is Hermitian. 
      Then, $A_k^* H_k + H_k A_k + 2\alphastar H_k = (M_k + \jmath N_k)^* (P_k+\jmath \Pi_k) + (P_k + \jmath \Pi_k) (M_k + \jmath N_k) + 2\alphastar(P_k + \jmath \Pi_k) = F + \jmath G \Lst0$, 
      with $F := M_k^\top P_k + P_k M_k + N_k^\top \Pi_k - \Pi_k N_k+2\alphastar P_k$ 
      and $G := M_k^\top \Pi_k + \Pi_k M_k - N_k^\top P_k + P_k N_k + 2\alphastar\Pi_k$. 
      On the other hand, it can be checked, after some computations, that $A_{e,k}^\top \smat{P_k  &  \Pi_k \\ -\Pi_k & P_k} + \smat{P_k  &  \Pi_k \\ -\Pi_k & P_k} A_{e,k} + 2\alphastar \smat{P_k  &  \Pi_k \\ -\Pi_k & P_k}= \smat{ F & G \\ -G & F}$, 
      which is negative definite, as to be proven, because 
      $\sigma\left(\smat{F & -G \\ G & F}\right) = \sigma(F+\jmath G)\cup\sigma(F-\jmath G)$,
      as shown in the proof of \ref{it:Ak_alpha} $\Longleftrightarrow$ \ref{it:Akreal_alpha}.

      For the discrete-time case, in view of Lemma \ref{lem:schur_complex}, Schur stability of $\frac{1}{\alphastar}A_k$ is equivalent to $A_k^* H_k A_k-(\alphastar)^2H \Lst 0$ for a positive definite Hermitian matrix $H_k=H_k^*$. By splitting real and imaginary part, we write $A_k = M_k + \jmath N_k$ and $H_k = P_k + \jmath \Pi_k$, where $P_k=P_k^\top$ and $\Pi_k = -\Pi_k^\top$ because $H_k$ is Hermitian.
      Then, 
      $A_k^* H_k A_k - {(\alphastar)^2}H_k = (M_k + \jmath N_k)^* (P_k+\jmath \Pi_k) (M_k + \jmath N_k) - {(\alphastar)^2}(P_k + \jmath \Pi_k) = F + \jmath G \Lst0$, with 
      $F := M_k^\top PM +N_k^\top PN - M_k^\top\Pi_k M_k + N_k^\top\Pi_k M_k-{(\alphastar)^2}P_k$ and 
      $G := M_k^\top \Pi_k N_k  - N_k^\top PM + M_k^\top\Pi_k M_k+N_k^\top\Pi_k N_k-{(\alphastar)^2}\Pi_k$.
      On the other hand, it can be checked, after some computations, that $ A_{e,k}^\top \smat{P_k  &  \Pi_k \\ -\Pi_k & P_k} A_{e,k} - {(\alphastar)^2}\smat{P_k  &  \Pi_k \\ -\Pi_k & P_k} = \smat{ F & G\\ -G & F}$, 
      which is negative definite, as to be proven, because 
      $\sigma\left(\smat{F & -G \\ G & F}\right) = \sigma(F+\jmath G)\cup\sigma(F-\jmath G)$,
      as shown in the proof of \ref{it:Ak_alpha} $\Longleftrightarrow$ \ref{it:Akreal_alpha}.
   
   \smallskip

   \noindent
   {\bf\em Proof of \ref{it:lyapReal_alpha} $\Longrightarrow$ \ref{it:lyap_alpha}}.
      According to Lemma~\ref{lem:L_frob}, there exists an orthogonal matrix  $T \in \myreal^{N \times N}$ 
      (satisfying $T^{\top}T = I_N $) 
      whose first column is $\frac{1}{\sqrt{N}} \ones$, and such that $\overbar L_d = T^\top L_d T $ is as in  \eqref{eq: L_bar_frob}.
      Let us now introduce the similarity transformation $ z = (T^{\top} \otimes I_{n}) x$. 
      Then, using the associative property of the Kronecker product, the dynamics in \eqref{eq:closedloop} can be rewritten as:
      \begin{eqnarray}
         \label{eq:intersys_changevariable}
         \dot z /z^+ &=& \overbar A z :=\left( (I_N\otimes A) - (\overbar L_d\otimes BC)\right) z
      \end{eqnarray}
      where the upper block-triangular structure of $\overbar L_d$
      carries over to matrix $\overbar A$, which can be written as
      \begin{equation}
         \label{eq:A_bar}
         \overbar{A} = \left[ 
         \begin{array}{c| c c c }
         A  & A_{12}  &\cdots & A_{1\nu}\\ \hline
         0 & F_1 & \cdots & A_{2\nu} \\
         \vdots & \vdots &  \ddots & \vdots\\
         0 &  0 & 0   & F_{\nu}
         \end{array}
      \right] = \left[
      \arraycolsep=1.3pt\def\arraystretch{1.3}
      \begin{array}{c| c  }
         A  & M_0\\ \hline
         \;\; 0  \;\;& \overbar A_1\\
      \end{array}
         \right] = \left[
      \arraycolsep=1.3pt\def\arraystretch{1.3}
      \begin{array}{c| c  }
         A  & M_0\\ \hline
         \;\; 0  \;\;& 
         \begin{array}{c| c  }
            F_1        & M_1\\ \hline
            \;\; 0 \;\;& 
            \overbar A_2 \\
         \end{array} \\
      \end{array}
      \right] = \left[
      \arraycolsep=1.3pt\def\arraystretch{1.3}
      \begin{array}{c| c}
      A  & M_0\\ \hline
      \;\; 0  \;\;& 
         \begin{array}{c| c  }
         F_1  & M_1\\ \hline
         \;\; 0  \;\;& 
            \begin{array}{c| c  }
            F_2  & M_2\\ \hline
            \;\; 0  \;\;& 
            \overbar A_3 \\
            \end{array} \\
         \end{array} \\
      \end{array}
      \right],
      \end{equation}
      and so on.
      In (\ref{eq:A_bar}) we introduced the following notation:
      \begin{align*}
         \overbar A_k &:= \begin{bmatrix}
         F_k & A_{k+1,k+2} & \cdots & A_{k+1, \nu}\\
         \;\; 0 \;\; & F_{k+1} & \cdots & A_{k+2,\nu}\\
         \;\; 0 \;\;& \;\; 0 \;\;& \ddots & \vdots \\
         \;\; 0 \;\;& \;\; 0 \;\;&\;\; 0 \;\; & F_{\nu}
         \end{bmatrix} ,\\
         M_{k-1} &:= \left[
         \begin{array}{cccc}
         A_{k, k+1}  & \quad& \dots & A_{k, \nu}
         \end{array}
         \right],
      \end{align*}
      for $k=1, \dots, \nu$ with $\overbar A_{\nu} = F_{\nu}$
      and where the matrices in the diagonal blocks are 
      induced by the structure of $\overbar L_d$ in \eqref{eq: L_bar_frob}. In particular,
      for each  $k=1, \dots, \nu$,
      if $\lambda_k \in \myreal$, then $F_k \in \myreal^{n\times n}$ is given by 
      $F_k = A_k = A -\lambda_kBC$. Then we may define $\overbar P_k := P_k$ 
      and obtain, 
      \begin{subequations} \label{eq:Lyap_block_real}
         \begin{align}
         \label{eq:Lyap_block_real_CT}
            F_k^\top \overbar P_k + \overbar P_k F_k \Lst - 2\alphastar \overbar P_k, \quad &\mbox{if }\Im(\lambda_k) = 0 \\
         \label{eq:Lyap_block_real_DT}
            \bigl(\mbox{resp.}\ \ F_k^\top \overbar P_k F_k \Lst {(\alphastar)^2} \overbar P_k, \quad & \mbox{if }\Im(\lambda_k) = 0 \bigr).
         \end{align}   
      \end{subequations}
      
      Instead, if $\lambda_k \in \complex$, then  
      $F_k \in \myreal^{2n\times 2n}$ is given by $F_k = (I_2\otimes A) - (\Lambda_k \otimes BC)$, 
      with the spectrum of $\Lambda_k$ being $\sigma(\Lambda_k) = \{ \lambda_k, \lambda_k^*\} = \{\alpha_k \pm \jmath \beta_k\}$, 
      namely there exists an invertible $S_k \in \myreal^{2\times 2}$ such that 
      $\Lambda_k  = S_k^{-1} \smat{\alpha_k & - \beta_k \\ \beta_k & \alpha_k}S_k$. Now, since $A_{e,k}$ in \eqref{eq:Aekdef} can be expressed as 
      $$ A_{e,k} = (I_2 \otimes A) - \left(\smat{\alpha_k & - \beta_k \\ \beta_k & \alpha_k} \otimes BC\right), $$
      then $F_k = (S_k^{-1} \otimes I_n) A_{e,k} (S_k \otimes I_n)$. As a consequence,
      from item~\ref{it:lyapReal_alpha},  
      using matrices  $\smat{P_k  &  \Pi_k \\ -\Pi_k & P_k}$ and noting that $\smat{P_k  &  \Pi_k \\ -\Pi_k & P_k} = (I_2\otimes P_k) + \left(\smat{0 & - 1 \\ 1 & O} \otimes \Pi_k\right)$,
      we may construct $\overbar P_k := (S_k^\top \otimes I_n) \smat{P_k  &  \Pi_k \\ -\Pi_k & P_k}  (S_k \otimes I_n) \Gst 0$, 
      so that \eqref{eq:Lyap_alpha_real} transforms into 
      \begin{subequations} \label{eq:Lyap_block_complex}
         \begin{align}
            \label{eq:Lyap_block_complex_CT}
               F_k^\top \overbar P_k + \overbar P_k F_k \Lst -  2\alphastar \overbar P_k, \quad &\mbox{if } \Im(\lambda_k) \neq 0 \\
            \label{eq:Lyap_block_complex_DT}
               \bigl(\mbox{resp.}\ \ F_k^\top \overbar P_kF_k \Lst  {(\alphastar)^2} \overbar P_k, \quad &\mbox{if } \Im(\lambda_k) \neq 0 \bigr).
            \end{align}
      \end{subequations}

      Based on \eqref{eq:Lyap_block_real} and \eqref{eq:Lyap_block_complex} and due to the block-triangular structure of $\overbar A_1$, there exist sufficiently small scalars $\overbar \eta_1, \ldots, \overbar \eta_{\nu-1}$ such that, defining $\breve P := \diag(\overbar \eta_1 \overbar P_1, \ldots,\overbar \eta_{\nu-1} \overbar P_{\nu-1}, \overbar P_{\nu})$ we have $\overbar A_1^\top \breve P + \breve P \overbar A_1 \Lst -2\alphastar \breve P$ (resp. $\overbar A_1^\top \breve P \overbar A_1 \Lst {(\alphastar)^2} \breve P$). 
      Additionally, we can always say that there exists $\alpha>\alphastar$ (resp. $0<\alpha<\alphastar$) such that $\overbar A_1^\top \breve P + \breve P \overbar A_1 \Leq -2\alpha \breve P$ (resp. $\overbar A_1^\top \breve P \overbar A_1 \Leq \alpha^2 \breve P$).
      As a consequence, we may partition $z = \smat{z_0 \\ \breve z}$, with $z_0 \in \mathbb{R}^n$ and $\breve z \in \mathbb{R}^{(N-1)n}$ and define function $W(z) := z^\top \overbar P z :=
      z^\top \diag(0, \breve P ) z =\breve z^\top \breve P \breve z$,
      which satisfies, along the trajectories of system \eqref{eq:intersys_changevariable},
      \begin{subequations} \label{eq:Wvar}
         \begin{align} 
         \label{eq:Wdot}
            \dot W(z) = \breve z^\top \left(\overbar A_1^\top  \breve P  +  \breve P  \overbar A_1 \right) \breve z  
            \leq -2\alpha\breve z^\top \breve P \breve z = -2\alpha W(z)\\
         \label{eq:Wplus}
            \bigl(\mbox{resp.}\ \ W(z^+)  = \breve z^\top \left(\overbar A_1^\top  \breve P \overbar A_1 \right) \breve z  
            \leq \alpha^2\breve z^\top \breve P \breve z = \alpha^2 W(z)\bigr). 
         \end{align}
      \end{subequations}
      Based on \eqref{eq:Wvar}, we define $V(x) := W((T^{\top} \otimes I_{n}) x)$ satisfying \eqref{eq:V_lyap_sandwich},\eqref{eq:V_lyap_var_CT} (resp. \eqref{eq:V_lyap_sandwich},\eqref{eq:V_lyap_var_DT}), which immediately implies \eqref{eq:V_lyap_var_CT} (resp. \eqref{eq:V_lyap_var_DT}), because $V$ corresponds to $W$ in the equivalent coordinates $x$. 

      To prove \eqref{eq:V_lyap_sandwich},
      let us first rewrite matrix $\overbar P := \diag(0, \breve P )$ as follows, 
      denoting by $e_1$ the first element of the Euclidean basis, and introducing 
      $\hat P :=\left( T \otimes I_n \right) \diag(I_N, \breve P )\left( T^\top \otimes I_n \right)\Gst0$, we have
      \begin{align}
         \nonumber
         \overbar P  &= \left( (I_N - e_1 e_1^\top) \otimes I_n \right) \diag(I_N, \breve P )
         \left( (I_N - e_1 e_1^\top) \otimes I_n \right) \\
         &= \left( (I_N - e_1 e_1^\top) T^\top \otimes I_n \right) \hat P
         \left( T (I_N - e_1 e_1^\top) \otimes I_n \right) .
         \label{eq:Pbar_expr}
      \end{align}

      Since the attractor $\A$ in \eqref{eq:setA} is a linear subspace generated by vectors $(\ones \otimes I_n)$,
      the distance $\abs{x}_{\A}$
      can be characterized in 
      terms of a suitable projection matrix $\Psi$ as follows:
         \begin{equation}
         \label{eq:distanceA}
         \abs{x}_{\A} = | \Psi x| = \left|
         \left(\left(I_N - \frac{1}{N}\ones \ones^\top\right) \otimes I_n \right) x 
         \right|,
      \end{equation}
      where $\Psi$ projects $x$ in the directions that are orthogonal to the generator of $\A$.

      Finally, using $W(z)= z^\top \overbar P z$, the identity in \eqref{eq:Pbar_expr},
      and noting that the orthogonality of $T$
      and its first column being $\frac{1}{\sqrt{N}} \ones$ 
      implies $T e_1 e_1^\top T^\top = \frac{1}{N} \ones \ones^\top$,
      we may express $V$ as
      \begin{align*}
      V(x) &=  x^\top (T \otimes I_{n}) \overbar P (T^{\top} \otimes I_{n}) x \\
      &= x^\top\left( (I_N - \tfrac{\ones \ones^\top}{N}) \otimes I_n\right) \hat P
      \left( (I_N - \tfrac{\ones \ones^\top}{N}) \otimes I_n \right)  x \\
      & = x^\top\Psi^\top \hat P \Psi  x ,
      \end{align*}
      where $\Psi$ is defined in \eqref{eq:distanceA}.
      The positive definiteness of $\hat P$ implies that $c_1 I_{N n} \Leq \hat P \Leq c_2 I_{N n}$ for some positive scalars $c_1,c_2$, which in turn implies
      \begin{align*}
         c_1 \left| \Psi x \right|^2 \leq V(x) \leq c_2 \left|\Psi x 
      \right|^2,
      \end{align*}
      thus implying \eqref{eq:V_lyap_sandwich}  by virtue of \eqref{eq:distanceA}.

   \smallskip

   \noindent
   {\bf\em Proof of \ref{it:lyap_alpha} $\Longrightarrow$ \ref{it:UGES_alpha}}. 
      For the continuous-time case, in view of the standard comparison lemma \citep[Lemma 3.4]{khalil3}, 
      condition \eqref{eq:V_lyap_var_CT} implies the existence of a
      uniform negative exponential bound on $V(x(t)) \leq {\rm e}^{-{2\alpha} t} V(x(0))$, along any solution $x$. This bound is easily extended to $|x|_{\A}^2$ using \eqref{eq:V_lyap_sandwich} in the following standard way:
      \begin{align*}
            |x(t)|_{\A}^2 \leq \frac{1}{c_1} V(x(t)) \leq 
            \frac{{\rm e}^{-{2\alpha} t}}{c_1} V(x(0)) \leq
            \frac{c_2}{c_1} {\rm e}^{-{2\alpha} t}   |x(0)|_{\A}^2.
      \end{align*}  
      
      For the discrete-time case, we can rewrite relation \eqref{eq:V_lyap_var_DT} as $V(x(t+1))\leq \alpha^2V(x(t))$.
      Then, taking advantage of the chain relation between consecutive time instant, we can state that $V(x(t))\leq\alpha^{2t}V(x(0))$.
      We can extend this bound to $|x|^2_\A$ using \eqref{eq:V_lyap_sandwich} in the following way:
      \begin{align*}
         |x(t)|_{\A}^2 \leq \frac{1}{c_1} V(x(t)) \leq 
         \frac{\alpha^{2t}}{c_1} V(x(0)) \leq
         \frac{c_2}{c_1}\alpha^{2t} |x(0)|_{\A}^2.
      \end{align*}

   \smallskip

   \noindent
   {\bf\em Proof of \ref{it:UGES_alpha} $\Longrightarrow$ \ref{it:attractor_alpha}}.
      From Lemma~\ref{lem:Laplacian_eig}, a zero eigenvalue of $L$ has non-negative left and right eigenvectors corresponding to $p$ and $\ones$.
      Non-negativity implies that the sum of the components of $p$ cannot be zero (otherwise the components of $p$ would all be zero and $p$ would not be an admissible eigenvector).
      Hence, $p^\top L = 0$ and $p^\top {\bf 1}_N \neq 0$.
      Consider then the dynamics of the 
      state $\tilde x_\circ(t):= \frac{1}{p^\top{\bf 1}_N}\sum\limits_{k=1}^{N} p_k x_k(t)$
      and note that from (\ref{eq:sys}):
      \begin{align*}
         \dot {\tilde x}_\circ (t) /{\tilde x}^+_\circ(t)
         = A \frac{1}{p^\top{\bf 1}_N}\sum\limits_{k=1}^{N} p_k x_k(t) + B \frac{1}{p^\top{\bf 1}_N}\sum\limits_{k=1}^{N} p_k u_k(t)
         = A \tilde x_\circ(t) - B \frac{1}{p^\top{\bf 1}_N} p^{\top}L  y = A \tilde x_\circ(t).
      \end{align*}
      Then $\tilde x_\circ$ evolves autonomously following \eqref{eq:xzero}, and corresponds
      to a linear combination of states $x_k$ weighted by the (non-negative) components of the eigenvector $p$.
      Uniform global $\alpha$--exponential stability of the attractor $\A$ implies that all states $x_k$ converge globally and exponentially to a common trajectory $\overbar x$ with convergence rate at least $\alpha$, 
      i.e., $\lim_{t \to +\infty} x_k(t) - \overbar x(t) = 0$ and $|x_k(t) - \overbar x(t)|\leq M\e^{-\alpha t}{|x_k(0) - \overbar x(0)|}$ (resp. $|x_k(t) - \overbar x(t)|\leq M\alpha^t |x_k(0) - \overbar x(0)|$) for all $k$. This common trajectory asymptotically coincides with the solution to the initial value problem \eqref{eq:xzero}, because, using $p^\top{\bf 1}_N =\sum_{k=1}^N p_k$, we obtain
      \begin{align*}
         \lim_{t \to +\infty} \tilde x_{\circ}(t) - \overbar x(t) 
         = \lim_{t \to +\infty} \frac{1}{p^\top \ones} \sum_{k=1}^N p_k x_k(t) - 
         \frac{\sum\limits_{k=1}^{N} p_k } {p^\top{\bf 1}_N}\overbar x(t)
         = \frac{1}{p^\top \ones} \sum_{k=1}^N p_k \lim_{t \to +\infty} ( x_k(t) - \overbar x(t)) = 0.
      \end{align*}
      
   \smallskip

   \noindent
   {\bf\em Proof of \ref{it:attractor_alpha} $\Longrightarrow$ \ref{it:Akreal_alpha}.}
      We prove this step by contradiction. 
      Therefore, we will prove that given a matrix $A_{e,k}+\alphastar I$ not Hurwitz (resp. a matrix $\frac{1}{\alphastar}A_{e,k}$ not Schur), the sub-states $x_i$ do not uniformly globally $\alphastar$--exponentially synchronize. In other words, we do not have synchronization to the unique solution of the following problem: 
      \begin{subequations} \label{eq:sync}
      \begin{align}
      \label{eq:sync_CT}
         \dot{\tilde {x}}_{\circ}=(A+\alphastar I){\tilde {x}}_{\circ} \\
      \label{eq:sync_DT}
         \left(\mbox{resp.}\ \ {\tilde {x}}_{\circ}^+  =\frac{1}{\alphastar}A{\tilde {x}}_{\circ}\right).
      \end{align}
      \end{subequations}
      Assume that one of the matrices $A_{e,k}$ in \eqref{eq:Aekdef} has spectral abscissa greater than $-\alphastar$ (resp. spectral radius grater than $\alphastar$), 
      and assume without loss of generality that it is $A_{e,\nu}$. 
      Consider the coordinate system in \eqref{eq:intersys_changevariable}
      with \eqref{eq:A_bar}. 
      Then, from the upper block-triangular structure of $\overbar{A}$ in \eqref{eq:A_bar}, we obtain that $\overbar A_{\nu}+\alphastar I$ is not Hurwitz (resp.  $\frac{1}{\alphastar}\overbar A_\nu$ not Schur) ($\overbar A_{\nu}=F_\nu$),
      and then there exists a vector $\omega^{*}\in \myreal^{n}$ (an eigenvector of one of the non-converging natural modes) such that the solution 
      to $\dot z=(\overbar A +\alphastar I)z$ (resp. $ z^+=\frac{1}{\alphastar}\overbar A z$)
      starting at $z^{*}(0)=\left[ 
         \begin{array}{c c c c}
         \zeros_n^{\top} & \dots & \zeros_n^{\top} & \omega^{*\top}
         \end{array} \right]^{\top}$ corresponds to
         $z^{*}(t)=\left[\begin{array}{c c c c}
         \zeros_n^{\top} & \cdots & \zeros_n^{\top} & z_{\nu}^{*\top}(t)
      \end{array}\right]^{\top}$, 
      where $z^*_{\nu}(t)$ does not converge to zero. 
      Define now a function $z \mapsto W(z) = \breve z^\top I_{N (n-1)} \breve z$
      as specified before \eqref{eq:Wvar}
      (that is, the same function $W$ for the specific selection $\breve P = I_{N (n-1)}$). 
      Evaluating $W$ along solutions, we get that 
      $W(z) =  |z^*_{\nu}(t)|^2$ does not converge to zero. 
      With $x \mapsto V(x)$ defined after \eqref{eq:Wvar}, we have that \eqref{eq:V_lyap_sandwich} holds,
      and  evaluating $V$ along the solution $x^*(t) = (T \otimes I_{n}) z^*(t)$, we have that $V$ does not converge to zero, 
      which implies that $x^*$ does not converge to the consensus set ${\A}$.
      In other words, 
      the components of $x^{*}(t)$ do not synchronize to the solution of \eqref{eq:sync} as to be proven.

\begin{remark}
   In many existing works, the Lyapunov function is a quadratic form similar to the construction of function $W$ just before equation \eqref{eq:Wdot}, but comprising the same blocks (of the form $I \otimes P_{\circ}$ for some suitable positive definite matrix $P_{\circ}$) on the side of the component $\breve{z}$ of vector $z$. Here, through the selection of matrix $\breve{P}$ comprising the matrices 
   $\overbar P_1, \ldots,\overbar P_{\nu}$ and the scaling factors
   $\overbar \eta_1 , \ldots,\overbar \eta_{\nu-1}$, we generalize that approach, possibly leading to exploitable degrees of freedom for optimized performance, as partially confirmed by the encouraging results of \citep{zaupa2023}.
\end{remark}

\section{Conclusions}
\label{sec:Conclusions}
   We have provided necessary and sufficient conditions for the synchronization of identical linear SISO systems, with a guaranteed convergence rate,
   both in the continuous-time and in the discrete-time case. Our conditions do not require any assumption on the graph, whose topology is just assumed to be time-invariant. Moreover, the proposed conditions also apply to systems where a direct input-to-output channel is present.
   While our conditions have been shown to be strategic for performance optimization in distributed feedback design in our preliminary work \citep{zaupa2023}, future work includes reinterpreting the main results of this paper for special classes of nonlinear systems, such as Lur'e systems with special sector-bounded nonlinearities. Our Lyapunov construction with non-identical diagonal blocks could also be exploited to reduce the conservatism of certain existing conditions addressing certain nonlinear synchronization problems.   

\section*{Acknowledgements}
   The authors thank Laura Dal Col for her contribution to a preliminary version of this paper, and Dimos Dimarogonas and Elena Panteley for useful discussions.

\bibliography{main_ejcon}

\end{document}